\documentstyle[12pt]{article}
\newfont{\feff}{cmti10}
\begin{document}

\begin{titlepage}
\title{
\begin{flushright}
{\bf\normalsize   PUPT-1332, revised}\\
\end{flushright}
$\frac{1}{N}$ Expansion and Particle Spectrum in Induced QCD}

\author{
A.A. Migdal\\
Physics Department,\\
Jadwin Hall, Princeton University\\
Princeton, NJ 08544-1000\\
}
\date{July 1992}

\maketitle

\begin{abstract}
We study the $ \frac{1}{N} $ expansion in the recently proposed model
of the lattice gauge theory induced by heavy scalar field in adjoint
representation. In the first approximation the fluctuations of the density
of eigenvalues of the scalar field are Gaussian, so that the scalar
glueball spectrum is defined from the corresponding linear wave
equation.

\end{abstract}
\end{titlepage}

\newpage

\section{Introduction}

The  problem of solving large $ N $ QCD seemed  hopeless until
recently some new approach was taken\cite{KazMig,Mig}. Instead of
trying to solve it starting from the asymptotic freedom, we suggested to
start much earlier, at such scales, that there are no gluons yet, but
rather some gluon constituents.

The idea is that one could then use the freedom of choosing any
constituents and any form of their interaction, as long as the
effective theory at large scales would be still QCD. Clearly the gauge
symmetry should be there from the start, but there need not be the
Yang-Mills (or Wilson, in the lattice formulation) term in the bare
Action. This term should be induced  in an asymptotically free region, as
result of integrating out heavy constituents.

This idea is quite old. Perhaps Sakharov was the first one to suggest
the induced Gravity. After that the idea was unsuccessfully
implemented by the author (see the review paper\cite{Mig83})  in the
fermionic string model of QCD. The
heavy constituents were introduced as 2-dimensional Majorana fermions
at the world sheet of string. Formally, the Schwinger-Dyson equations
of the large $N$ QCD were satisfied, but one could not make any sense
out of this model.

As we see it now, after ten more years of study of the string theory,
the real problem comes from the string. Apparently, there is no such
thing as a noncritical string in any number $ D > 1 $ of external
dimensions. The interaction at the world sheet must be essentially
nonlocal, and/or it should involve some extrinsic geometry.
The idea of string realization of the Lorentz group was a beautiful
one, and it still has chance to work in Gravity, but in QCD I
personally do not believe in this idea any more. Apparently, the
coordinates could only be arguments of fields, rather then fields in
QCD.

Coming back to the induced QCD, the idea of Kazakov and myself was to
take heavy scalar field in adjoint representation of the gauge group
as a constituent field.
Naively, if one just integrates this field out, one induces the
correct Yang-Mills action, with positive coefficient in front, scaling
correctly at large $N$. By the way, the simpler choice of the
fundamental representation would yield too large induced gauge
coupling, so it could not induce an asymptotically free QCD at large N.
The large ($ \sim N $) number of fundamental scalars would do it, but
this model is hard to solve.

The correct computation of induced gauge coupling is not so trivial,
as there are feedback effects from induced hard gluons, and  those from
effective scalar quartic interaction.
If one starts from asymptotically free region, and goes to smaller
scales, there would be two conflicting effects. First, there would be
antiscreening of gauge coupling from the gluons, which would tend to
decrease the gauge coupling further. Second, there would be screening
from the scalars, which is numerically smaller, then antiscreening, so
that alone it cannot stop the logarithmic decrease of effective gauge
coupling. The desired infinite gauge coupling at lattice scales could
be generated by  scalar quartic interaction, which is known to grow
without limits at small scales. At least we know, that asymptotic
freedom is not consistent with scalar quartic interactions. At larger
scales, the scalars' decouple, being heavy, so that true asymptotic
freedom is left intact. So, the scalars effectively cut off the
asymptotic freedom at the scale of their mass, and move the effective
coupling towards another "confinement" region at lattice scales.

What happens in this model at small scales is  beyond the
reach of the perturbative field theory. However, if we take the particular
nonperturbative theory, we have a chance to adjust the parameters so,
as to visit the perturbative region on the way from small to
physical scales. In fact, it is hard to imagine how could
we miss the asymptotic freedom, if we would get nontrivial theory in
four dimensions. The common belief is that there is only one gauge
theory in four dimensions: the asymptotically free, quark confining QCD.

The unique property of the scalar field in adjoint representation is
that corresponding lattice gauge model without the Wilson term is soluble
at large $N$, for arbitrary scalar potential. The one link integrals\cite{IZ}
over the gauge fields $ U_{\mu} $ reduce at large $N$ to certain functional
determinant, which was analyzed in previous works\cite{KazMig,Mig}. In
particular, in the second work\cite{Mig}, we found some coupled
integral equations, exactly soluble by means of the Riemann-Hilbert
method.

The gauge field integration produces effective action, which depends
only on the density of eigenvalues of the scalar field. One could
therefore, integrate out the unitary matrices, which diagonalize the
scalar field, or simply take the gauge where this field is diagonal.
The corresponding Faddeev-Popov Jacobian is nothing but the
Vandermonde determinant, which produces in effective Action the term,
quadratic in the density of eigenvalues.

At $ N= \infty $   this density does not fluctuate: it is given by the
solution of the classical equations, minimizing this effective Action.
By some historical reasons it was difficult for everybody, authors
included, to accept this simple and natural scenario. We were misled
by large $N$ matrix models of 2D Quantum Gravity, where the
Itzykson-Zuber determinant was used in a different context. There, it
reduced to the Slater determinant, thus revealing the
hidden fermionic degrees of freedom. In higher dimensions there is no
such thing as fermionization of the Bosonic problem, neither could there
be a spatially inhomogeneous vacuum. So, there seems to be no
alternative to the space independent vacuum density of the eigenvalues of
the scalar field, which we call the Master Field.

Still, the technical problem of reexamining the old $C=1$ matrix model
with the new representation of the Itzykson-Zuber determinant was
pending, so  the author (unpublished) checked within the strong
coupling expansion, that the old fermionic solution can also be
correctly reproduced by the spatially independent Master Field.
Recently, this problem was completely solved\cite{Gross}, so that
there is no longer any contradiction between the fermions and the
Master Field.

The classical solutions for the density come in two phases: the
strong coupling phase, with density continuous at the origin, and the
weak coupling phase, where there is a finite gap in the spectrum. The
critical phenomena near the transition point are much richer then in
the usual one-matrix models. The critical indices, computed
in\cite{Mig} are transcendental numbers at $ D> 2 $, in particular,
the density grows as $ \rho(\lambda) \sim \lambda^{\alpha} $ and
physical mass scale $ m^2 \sim \lambda^{\alpha-1} $ where $ \cos \pi
\alpha = -\frac{D}{3D-2} ,  \alpha > 1$.

The scaling laws in the four-dimensional theory are amazing, but they do
not contradict the asymptotic freedom, as there is extra
renormalization from the lattice scales to the region of the asymptotic
freedom. In fact, the induced gauge coupling $ \beta_0 =
\frac{1}{Ng_0^2} $ is expected to come out as some negative number
times logarithm of the scalar mass in the lattice units, according to
the RG analysis at the small scale region. Comparing this with the
usual RG analysis in the large scale region, we, indeed, get the
scaling laws for the physical mass in terms of the bare scalar mass.

There is, however, one unsolved problem in the induced QCD, that of the
spurious local Abelian symmetry. It is well known, that the lattice
theory in adjoint representation has extra local $Z_{N}$ symmetry.
This issue was analyzed a long time ago\cite{Khokh,Mak}, and it was noted,
that unbroken symmetry locally confines quarks so that they cannot
move separately even at the lattice scales. It was suggested in the
recent work\cite{Kogan}, that this symmetry breaks spontaneously at
the transition point, as it does in the mean field
approximation\cite{Khokh,Mak}.
In any case, this is the problem for the higher $  \frac{1}{ N} $
approximations, as the center of the gauge group is negligible at $ N
= \infty $.

The present paper is devoted to the next $  \frac{1}{ N} $ correction
in the scalar sector of the strong coupling phase of the induced QCD.

In Section 2 we develop the general method of $  \frac{1}{ N } $
expansion, and find exact integral equations for the Gaussian
fluctuations of the density.

In Section 3 we go to the local limit, using the same method, as
before, and we find another Riemann-Hilbert problem, soluble in the
scaling limit.

In the Appendix we present the computation of the second variation of the
Itzykson-Zuber determinant.

\section{$\frac{1}{N}$ Expansion}

At infinite $ N $ there are no fluctuations of the density of
eigenvalues of the scalar field, and therefore, no interesting
physics. These fluctuations show up in the first order of the $
\frac{1}{N} $ expansion. There are two methods for the systematic $
\frac{1}{N} $ expansion: the Schwinger-Dyson equations and the saddle
point method.

For our present purposes the saddle point method appears to work better,
so we elaborate it here, using the modification suggested by G.Parisi
(unpublished). The idea is to change the integration variables from
the eigenvalues to the corresponding density
\begin{equation}
  \rho(\lambda)=  \frac{1}{N} \sum_j \delta(\lambda-\lambda_j)
\end{equation}
The change of variables goes as follows
\begin{equation}
  \int d^N \lambda \propto \int {\cal D} \rho J[\rho]
\end{equation}
where
\begin{equation}
  J[\rho] = \int {\cal D}\epsilon\exp
	\left(
 \imath N \int d \lambda
\epsilon(\lambda) \rho(\lambda)
	\right)
\int d^N \lambda \exp
	\left(
	 -\imath \sum_j \epsilon(\lambda_j)
	\right)
\end{equation}
We neglect
everywhere the constant normalization factors, as we are interested only in
averages.

It is  convenient to shift $ \epsilon(\lambda) $ by $ \imath \ln
\bar{\rho}(\lambda) $ , and then single out the translational
mode $ \epsilon = const $ which yields the density normalization condition
\begin{equation}
  \delta
\left(1-\int d \lambda \rho(\lambda) \right)
\end{equation}
The translational mode can be eliminated by the background gauge condition
\begin{equation}
  \delta \left(\int d \lambda
\bar{\rho}(\lambda) \epsilon(\lambda) \right)
\end{equation}
with the trivial constant Jacobian, which does not require ghosts.
After simple transformations we find
\begin{equation}
  J[\rho] = \delta
\left(1-\int d \lambda \rho(\lambda) \right)
\exp \left(- N \int d \lambda \rho(\lambda)\ln
\bar{\rho}(\lambda) \right) \tilde{J}[\rho]
\end{equation}
where
\begin{eqnarray}
\tilde{J}[\rho] &=&
\int {\cal D}\epsilon \delta \left(\int d \lambda
\bar{\rho}(\lambda) \epsilon(\lambda) \right)
\exp \left(-\frac{N}{2} \int d \lambda
\bar{\rho}(\lambda)\epsilon^2(\lambda) \right) \\ \nonumber & \,&
\exp N \left(
 \imath  \int d \lambda
\epsilon(\lambda) \rho(\lambda) + \frac{1}{2}  \int d \lambda
\bar{\rho}(\lambda)\epsilon^2(\lambda)
+ \ln
\left(
\int d \lambda\bar{\rho}(\lambda)
\left(\exp(-\imath \epsilon(\lambda)) \right)
	\right) \right)
\end{eqnarray}

Now we could expand the last exponent in $ \epsilon $, keeping in
mind, that density fluctuations $ \rho-\bar{\rho}\sim \frac{1}{ N} $
and $ \epsilon\sim N^{-\frac{1}{2}} $. The resulting expansion of $
\tilde{J} $ is
straightforward, this is one dimensional perturbation theory with
propagator
\begin{equation}
  \left\langle \epsilon(\lambda) \epsilon(\lambda') \right\rangle  =
 \frac{1}{ N} \left(\frac{\delta(\lambda-\lambda')}{
\bar{\rho}(\lambda)} -1 \right)
\end{equation}
In the leading order we could set $ \tilde{J} = 1 + O[
\frac{1}{N}] $, which leaves us with extra term
\begin{equation}
    N \int d \lambda \rho(\lambda)\ln
\bar{\rho}(\lambda)
\end{equation}
in effective Action.
This term is local and linear in density, so it effectively shifts the scalar
field potential, without altering the quadratic form of the second
variations.

This quadratic form can be computed exactly, using the technique of
the previous work\cite{Mig}. Let us outline this computation. First let us
rederive the saddle point equation with the new technique.
The normalization condition would be satisfied identically for
\begin{equation}
  \rho(\lambda) = \bar{\rho}(\lambda) -\frac{\psi'(\lambda)}{N}
\end{equation}
where $ \psi(\lambda) $ vanishes at the endpoints of the support of $
\bar{\rho}(\lambda)$.
The first variation with respect to $ \psi(\lambda) $ is equivalent to
differentiation $ \frac{d}{d \lambda_i}$ and setting $ \lambda_i
\rightarrow \lambda $, as one can readily check.

The effective Action for the $x$ - dependent density $\rho_x $ reads
\begin{eqnarray}
  \lefteqn{S_{eff}[\rho] =}\\ \nonumber & \,&
 -N^2 \sum_x \int d \lambda \rho_x(\lambda) \int d \lambda'
\rho_x(\lambda') \ln| \lambda-\lambda'| + N^2 \sum_x\int d \lambda
\rho_x(\lambda)  U(\lambda)
- \sum_{<xy>} \ln I[\rho_x,\rho_y] \\ \nonumber & \,&
+   \sum_x \left(N\int d \lambda
\rho_x(\lambda)\ln \bar{\rho}(\lambda) + \ln \tilde{J} \right)
\end{eqnarray}
where the first term comes from the Vandermonde determinant, the
second one from the scalar field potential, the third one from the
Itzykson-Zuber determinant, and the remaining terms - from the above
Jacobian. These last terms would contribute only in the higher orders of
the $  \frac{1}{ N} $ expansion.

The background density of eigenvalues \footnote{ In the rest of the
paper we denote the background density as $ \rho $ rather then
$\bar{\rho}$.} is determined at $ N= \infty$ from the saddle point equation
\begin{equation}
  \frac{\delta S_{eff}[\rho]}{\delta \psi(\lambda)} =  \frac{1}{ N}
\frac{d}{d \lambda} \frac{\delta S_{eff}[\rho]}{\delta \rho(\lambda)}  =0
\end{equation}
This equation is the same, as before\cite{Mig}
\begin{equation}
  F(\lambda) = \frac{-2 \Re\, V'(\lambda)+U'(\lambda)}{2 D}
\end{equation}
where
\begin{equation}
  V'(\lambda) = \int d \lambda'  \frac{\rho(\lambda')}{\lambda-\lambda'}
\end{equation}
and the function $ F(\lambda) $ represents the logarithmic derivative
of the Itzykson-Zuber determinant.
This function for arbitrary density satisfies the set of
Schwinger-Dyson equations, which were derived and studied in the
previous work.  Substituting the  classical formula for $
F(\lambda)$ in these equations, we arrived\cite{Mig} at the Master
Field Equation (MFE)
\begin{equation}
  R(\lambda) = \frac{1}{2D}U'(\lambda) + \frac{D-1}{D} \Re \,V'(\lambda)
\end{equation}
\begin{equation}
  \Re\,V'(\lambda)= \int\frac{ d \lambda'}{\pi}\arctan \frac{\pi
\rho(\lambda')}{\lambda-R(\lambda')}
\label{MFE}
\end{equation}

Our objective now is to compute the second variation of the effective
Action. The potential term does not contribute, being linear in
density, the Vandermonde determinant yields a local term
\begin{equation}
 -\wp\, \int d \lambda \psi(\lambda)\int d\lambda'\psi(\lambda')
\frac{1} {(\lambda-\lambda')^2}.
\end{equation}
The Itzykson--Zuber determinant yields two terms, the local one being
\begin{equation}
  -D\int d \lambda \psi(\lambda) \int d\lambda'\psi(\lambda')
\sigma(\lambda,\lambda')
\end{equation}
with
\begin{equation}
\sigma(\lambda,\lambda') =
\frac{d}{d \lambda'} \frac{\delta F(\lambda)}{\delta \rho(\lambda')}
\end{equation}
and the nonlocal one
\begin{equation}
  - \frac{1}{2}\int d \lambda \psi_x(\lambda)\int d \lambda'
\sum_{\mu=-D}^{D} \psi_{x+\mu}(\lambda') \eta(\lambda,\lambda')
\end{equation}
with
\begin{equation}
   \eta(\lambda,\lambda')= \frac{d}{d \lambda'} \frac{\delta
F(\lambda)}{\delta \rho_{x+\mu}(\lambda')}
\end{equation}

Collecting all the finite terms, and going to the local limit, we find
the standard form of effective action of continuum field theory
\begin{equation}
S_{eff} =\int d^D x \,{\cal L}(x) 
\end{equation}
with effective Lagrangean
\begin{eqnarray}
& & \lefteqn{{\cal L }(x) = \int d \lambda  \int d\lambda'\,
 \frac{1}{2}\eta(\lambda,\lambda')\, \partial_{\mu} \psi(x,\lambda)
\,\partial_{\mu} \psi(x,\lambda')} \\ \nonumber & & -
\left(\frac{1}{(\lambda'-\lambda)^2} +\,
D \left(\sigma(\lambda,\lambda') + \,\eta(\lambda,\lambda') \right)
\right) \psi(x,\lambda)\,\psi(x,\lambda') \\ \nonumber & \,&
+ \int d \lambda \psi(x,\lambda)\frac{\rho'(\lambda)}{\rho(\lambda)}
\label{S2Continuum}
\end{eqnarray}

The linear term shifts the vacuum average of the density by
\begin{equation}
  \delta \rho(\lambda) = -  \frac{1}{ N}\psi_0'(\lambda)
\end{equation}
where $\psi_0$ satisfies the equation
\begin{equation}
  \int d \lambda'\,\psi_0(\lambda')\left(\frac{1}{(\lambda'-\lambda)^2} +\,
D \left(\sigma(\lambda,\lambda') + \,\eta(\lambda,\lambda') \right)
\right) = \frac{\rho'(\lambda)}{2 \rho(\lambda)}
\end{equation}

The particle spectrum is described by the wave equation for the
fluctuations $ \psi - \psi_0 = Y(\lambda) e^{\imath P x} $
\begin{equation}
 0= \int d\lambda'\,Y(\lambda')
 \left(\frac{1}{(\lambda'-\lambda)^2} +\,
D\sigma(\lambda,\lambda') + \,(D-\frac{1}{2}
P^2)\eta(\lambda,\lambda') \right)
\end{equation}

In the Appendix we study the perturbed Schwinger-Dyson equation for the
second variations and we find the following linear integral equations
for the kernels $ \sigma, \eta $
\begin{equation}
  \int d \lambda'\, {\cal K}(\lambda_0,\lambda')\eta(\lambda',\lambda'') =
\frac{1}{(\lambda_0-\lambda'')^2} \label{Kernel}
\end{equation}
\begin{equation}
\int d \lambda'\, {\cal K}(\lambda_0,\lambda') \sigma(\lambda',\lambda'') =
\int d \lambda'\, {\cal K}(\lambda_0,\lambda')
	\left(
	- \frac{1}{ (\lambda'-\lambda'')^2}+  \frac{1}{
G(\lambda_0,\lambda')} \frac{d}{d
\lambda''}\frac{G(\lambda_0,\lambda'')}{\lambda'-\lambda''}
	\right)
\end{equation}
where
\begin{equation}
  {\cal K}(\lambda_0,\lambda') =
\frac{\rho(\lambda')}{(\lambda_0-R(\lambda'))^2 + \pi^2 \rho^2(\lambda')}
\end{equation}
and
\begin{equation}
  G(\lambda_0,\lambda') =  \frac{1}{ \lambda_0-R(\lambda')} \Re\, \exp \left(
\int\frac{ d \lambda''}{\pi(\lambda''-\lambda')}\arctan \frac{\pi
\rho(\lambda'')}{\lambda_0-R(\lambda'')} \right)
\end{equation}

Now we  eliminate the unknown functions $ \sigma, \eta $ by integrating
the wave equation over $ \lambda $ with the weight $ {\cal
K}(\lambda_0,\lambda) $ which yields
\begin{eqnarray}
 &\, & D \int
\frac{d\lambda'\,\psi_0(\lambda')}{(\lambda_0-\lambda')^2} -
\int\int d \lambda\,d\lambda'\,\psi_0(\lambda') {\cal K}(\lambda_0,\lambda)
	\left(
	\frac{D-1}{(\lambda-\lambda')^2}- \frac{D}{
G(\lambda_0,\lambda)} \frac{d}{d
\lambda'}\frac{G(\lambda_0,\lambda')}{\lambda-\lambda'}
	\right) \\ \nonumber & \,&
= \int d \lambda {\cal K}(\lambda_0,\lambda)
\frac{\rho'(\lambda)}{2 \rho(\lambda)}
\label{Psi0}
\end{eqnarray}
\begin{equation}
  (D-\frac{1}{2} P^2) \int
\frac{d\lambda'\,Y(\lambda')}{(\lambda_0-\lambda')^2} =
\int\int d \lambda\,d\lambda'\,Y(\lambda') {\cal K}(\lambda_0,\lambda)
	\left(
	\frac{D-1}{(\lambda-\lambda')^2}- \frac{D}{
G(\lambda_0,\lambda)} \frac{d}{d
\lambda'}\frac{G(\lambda_0,\lambda')}{\lambda-\lambda'}
	\right)
\label{Wave}
\end{equation}
These equations should be solved together with the MFE.

All above equations were exact in a sense that we did not go to the
local limit, except for replacing the lattice derivatives by local
ones \footnote{One could restore the lattice theory by replacing $ \frac{1}{2}
P^2$ by $ 2 \sum_{\mu} \sin^2 \frac{1}{2} P_{\mu}  $.}, neither did we
assume anything about the support of eigenvalues. The same equations
hold for the weak coupling phase, where there is a gap at the
origin. In this paper we study the strong coupling phase
where there is no such gap.

\section{Local Wave Equation}

Let us consider the local limit, when the density at the origin
vanishes. The analysis of the Master Field Equation\cite{Mig} shows,
that the scaling solution, independent of the ultraviolet cutoff,
requires, that $ \rho(\lambda) $ and $ r(\lambda) = R(\lambda) -\lambda $
vanish as some power of the cutoff, so that
\begin{equation}
  \rho(\nu) \sim r(\nu) \ll \nu
\end{equation}
In this case the MFE can be expanded in $ \rho,r $ as follows
\begin{eqnarray}
& & \left(\rho(\lambda )r(\lambda )\right)''  = \mbox{regular terms}-
\wp  \int_{- \infty}^{ \infty} d\nu \,
\frac{\rho^2(\nu)}{(\lambda -\nu)^3} \\
\nonumber & \,&
r(\lambda ) = \frac{1}{2D} u'(\lambda )+\frac{D-1}{D} \Re\,V'(\lambda)
\label{scaling}
\end{eqnarray}
where renormalized potential $ u'(\nu) $ starts from the linear mass term
$ m^2 \nu $. The higher order terms are irrelevant in the scaling
region, but they might be important for the full MFE, to provide
required cancellations of the regular terms in the scaling region.
These regular terms represent substruction terms in dispersion
relations. There is no need for these terms in the full MFE at the
lattice, but in the local theory there are ultraviolet divergencies.

This convenient method of elimination of regular terms is to
introduce two analytic  functions
\begin{equation}
  {\cal P}(z) = \frac{u'(z)}{2(1-D)}-V'(z)
\end{equation}
\begin{equation}
  {\cal Q}(z) =  \mbox{regular terms}
+ \pi \int_{- \infty}^{ \infty} d\mu \,\frac{\rho^2(\mu)}{\mu-z}
\end{equation}
with the symmetry property
\begin{equation}
  {\cal P}(-\bar{z}) = -\bar{{\cal P}}(z)\\;\;
  {\cal Q}(-\bar{z}) = -\bar{{\cal Q}}(z)
\end{equation}
and note that at $ \Im z \rightarrow +0$ by construction
\begin{equation}
  \Im\,{\cal Q} =  \left(\Im \,{\cal P}\right)^2.
\end{equation}
On the other hand, in virtue of the  above equation for density
\begin{equation}
  \Re \, {\cal Q} = \frac{1-D}{D} \Im \,\left({\cal P}^2\right)
\end{equation}
This  is the nonlinear Riemann-Hilbert problem.

The key identity, used in derivation of  local MFE  is
the following one
\begin{equation}
  \int d \lambda' {\cal K}(\lambda,\lambda')A(\lambda') \rightarrow
A(\lambda) -(r(\lambda)A(\lambda))' + \wp\, \int \frac{d
\lambda'\rho(\lambda')A(\lambda')}{(\lambda-\lambda')^2} + O(\rho^2)
\end{equation}
where the first two terms come from the small region $ \lambda' -
\lambda \sim \rho(\lambda)$, and the last one comes from the region $
\lambda' \sim \lambda $. The integral in the small region can be
reduced to the residue at the complex pole of the kernel.
To be more precise, the $ O(\rho^2) $ correction to this formula was also used
in
derivation of the local MFE, where the linear terms cancel, but we do
not need this correction below.

The  expansion of the function $ G(\lambda,\phi) $ up to linear terms
in $ \rho $ reads
\begin{equation}
  G(\lambda,\phi) \rightarrow  \frac{1}{(\lambda-\phi-r(\phi))} \left(
1+ \wp\, \int \frac{d \nu \rho(\nu)}{(\phi-\nu)(\lambda-\nu)} \right)
\end{equation}
Substituting this into the wave equation (\ref{Wave}), we find variety of
terms. To reduce them, it is convenient to introduce the analytic function
\begin{equation}
  {\cal F}(z) = \int \frac{d \phi \rho(\phi)}{\phi-z} \, \wp \int
\frac{d \phi' Y(\phi')}{(\phi-\phi')^2}
\end{equation}
such that
\begin{equation}
  \Im\, {\cal F}(\phi+ \imath 0) = \pi \rho(\phi) \, \wp \int
\frac{d \phi' Y(\phi')}{(\phi-\phi')^2}
\end{equation}
The function $ Y(\phi)$ can be reconstructed from dispersion relation
\begin{equation}
  \wp\,\int d \phi' Y'(\phi') \frac{1}{\phi'-\phi} =
\frac{\Im\, {\cal F}(\phi+ \imath 0)}{\pi \rho(\phi)}
\end{equation}
\begin{equation}
  Y(\phi) = \int d\nu \frac{\Im\, {\cal F}(\nu+ \imath
0)}{\pi^3 \rho(\nu)} \ln |\phi-\nu|
\end{equation}

Let us now turn to the wave equation.
After some algebra, we reduce it  to the following
boundary problem\footnote{ The real and imaginary parts at real $ \phi
$ are understood, as usual limits from the upper half plane.}
\begin{equation}
  2 \pi \rho(\phi)\Re {\cal F}'(\phi)+
\left(-P^2 + 2 +2 D \Re\,V''(\phi)\right)\,\Im {\cal F}(\phi) +
2(D-1)\rho(\phi)\left(\frac{r(\phi)\,\Im {\cal F}(\phi)}{\rho(\phi)} \right)'
=0
\end{equation}

Let us now solve these problems in the scaling limit.
The local MFE allows the scaling solution, in proper units
\begin{equation}
 \pi \rho(\lambda) = \lambda^{\alpha} \\;\;
\alpha= 1+  \frac{1}{\pi} \arccos \frac{D}{3D-2}
\end{equation}
Various branches of the arccosine correspond to various fixed points
of the theory. This scaling form holds for $ \lambda \gg m^{\gamma}$  where
$ \gamma=\frac{2}{\alpha-1} $, and $ m$ is the physical mass scale.

In this case the functions $ V'(\lambda),r(\lambda) $ are
given by
\begin{equation}
  \Re\, V'(\lambda) \rightarrow -\frac{m_1^2+2}{2D} \lambda -
\lambda^{\alpha} \tan \frac{1}{2} \pi \alpha
\end{equation}
\begin{equation}
  r(\lambda) \rightarrow  - \lambda^{\alpha}
\frac{D-1}{D}  \tan \frac{1}{2} \pi \alpha
\end{equation}
The relation between the mass $ m_{1}^2 $ and the
scalar potential is as follows
\begin{equation}
  m_0^2 \equiv U''(0) = 2 D + u''(0) = 2 D + 2 -\frac{2}{D} +
\frac{D-1}{D} m_1^2
\end{equation}
so that the critical point is at
\begin{equation}
  m_{0,c}^2 = 2 \frac{D^2 + D-1}{D}
\end{equation}

In this paper we shall find the solution at $ \phi \gg m_{1}^{\gamma} $. Then
we are left with the scaling
terms, so that the scaling Ansatz for $ {\cal F}$ could be used
\begin{equation}
  {\cal F}(z) = \imath^s (-\imath z)^{\epsilon}\\;\; s=\{0,1\}
\end{equation}
The discrete parameter $s$ corresponds to the parity
\begin{equation}
  {\cal F}(-\bar{z}) = (-1)^s \bar{{\cal F}}(z)
\end{equation}
and the ratio of real and imaginary parts follows from symmetry and
analyticity.
Substituting this Ansatz into the equation, we find the equation for
the index $ \epsilon $
\begin{equation}
 0=g_{s}(\epsilon)\equiv \tan \left(\frac{\pi \alpha}{2} \right) \left(D
\alpha+\frac{(D-1)^2}{D} \epsilon \right) + \epsilon \cot  \left(\frac{\pi
(\epsilon -s)}{2}\right)
\end{equation}
There are infinitely many solutions $ \epsilon_{s,n} $ which grow
asymptotically linearly with the principal quantum number $ n $.

In general, there are also powerlike corrections to the
density, so that the solution is not so elementary.  The complete
computation of the spectrum
requires an exact  solution of the nonlinear Riemann-Hilbert
problem for the density. This is the subject of the next paper.

One important thing is clear right away: this is not a free field
theory at $ D>1 $, neither we see any pathology at $ D>1 $, like those
of the Liuoville theory. This is confining theory, but not quite the
string. In general, there are infinitely many masses, and they grow
 with quantum numbers, as one would expect for QCD.

\section{Discussion}

The puzzle still remains unsolved. Is this QCD?

At least this is the theory of scalar particles, confined by strong
interaction with gluon field. Would gluon field decouple, like it does
in one dimension, we would get trivial free particle spectrum.
Instead, we are going to get infinitely many  particles, with growing masses,
stable at infinite $ N $. We expect the decays to show up in the next
$  \frac{1}{ N } $ approximations.

In QCD, we would like to get the mesons, from confined $ \bar{q} q $
pair, in addition to pure glueballs. However, the quarks are locally
confined in this model, unless the $Z_N $ symmetry would break,
spontaneously or otherwise. This seems to be the most urgent thing to
do.

Also, we would like to see the perturbative QCD. This can be checked
already in the scalar sector. If there is admixture of glueballs to
the scalar branch of the spectrum, then the correlation function of
fluctuations of the scalar eigenvalue density should have logarithmic
singularity at $ P^2 \rightarrow \infty $, times some power of $ P^2
$. This singularity in perturbative QCD  comes from the two gluon
exchange, via the $ F_{\mu\nu}^2 $ operator. Here it must come out
as something like
\begin{equation}
  \sum_n  \frac{1}{ P^2 + \mbox{const } n} \sim \ln P^2
\end{equation}
due to the divergencies of the spectral sums.

Everything seems to be set up for the comparison of induced QCD with
perturbative QCD. Let us work hard and keep the fingers crossed.

\section{Acknowledgements}

I am grateful to Giorgio Parisi  for  useful suggestions, and to all
members of the theory group in University 2 of Rome, where this work
was completed, for friendly and  creative atmosphere.
This work was partially supported by the National Science Foundation under
contract PHYS-90-21984.

\appendix

\section{Second Variation of Itzykson-Zuber Integral}

In the previous paper we  derived  the following coupled set of linear
integral equations  for functions $ G_{\lambda}(\nu),F(\nu) $
\footnote{ In this Appendix we treat $ \lambda $ as subscript rather
as argument of $ G $.}
\begin{equation}
\int d \mu \rho_1(\mu)
\frac{G_{\lambda}(\mu)-G_{\lambda}(\nu)}{\mu- \nu} = -1 + (\lambda
-F(\nu)) G_{\lambda}(\nu) \label{Gequation}
\end{equation}
\begin{equation}
\int d\mu \rho_1(\mu) G_{\lambda}(\mu) =  \int d\nu
\frac{\rho_2(\nu)}{\lambda -\nu}
\label{Extra}
\end{equation}
where $ F(\nu) $ is the first variation of the logarithm of the
Itzykson-Zuber integral (see the section 2.). The densities of
eigenvalues at two endpoints of the link were denoted as $ \rho_1
$ and $ \rho_2 $. In order to find the second variations, we
should vary this equation with respect to $ \rho_1 $ and $
\rho_2 $.

Let us compute nonlocal variation $ \eta$ first, as
it is technically simpler. In this Appendix we denote the eigenvalues
as $ \phi , \mu $ and $ \lambda $.  We are
going to vary $ F, G $ with respect to $ \rho_2 $. These variations
\begin{equation}
\eta(\psi ,\phi)= \frac{d}{d \psi} \frac{\delta
F(\phi)}{\delta \rho_2(\psi)}\\;\;
g_{\lambda}(\psi,\phi) = \frac{d}{d \psi} \frac{\delta
G_{\lambda}(\phi)}{\delta \rho_2(\psi)}
\end{equation}
satisfy the differentiated version of above equations
(\ref{Gequation}), (\ref{Extra})
\begin{equation}
 \eta(\psi,\nu)G_{\lambda}(\nu)=(\lambda -F(\nu)) g_{\lambda}(\psi,\nu) -
\int d \mu \rho(\mu)
\frac{g_{\lambda}(\psi,\mu)-g_{\lambda}(\psi,\nu)}{\mu- \nu}
 \label{gequation}
\end{equation}
\begin{equation}
\int d\mu \rho(\mu) g_{\lambda}(\psi,\mu) =
\frac{1}{(\lambda-\psi)^2} \label{extra}
\end{equation}

As in\cite{Mig}, we have to introduce the auxiliary analytic function
\begin{equation}
{\cal A}_{\lambda,\psi }(z) = \int d \mu \frac{\rho(\mu)
g_{\lambda}(\psi,\mu)}{z- \mu} 
\end{equation}
in the upper half plane of $ z $ and we find from above equations the following
linear relations for the corresponding boundary values
\begin{equation}
{\cal A}_{\lambda,\psi }(z+\imath 0)\, \left(\lambda -R(z)- \imath\pi
\rho(z) \right) -
{\cal A}_{\lambda,b}(z-\imath 0)\, \left(\lambda -R(z) +\imath\pi
\rho(z) \right) = - 2
\pi \imath \rho(z) G_{\lambda}(z) \eta(\psi,z) 
\end{equation}
At infinity this function decreases as $\frac{1}{z} $ with coefficient
in front known from (\ref{extra})
\begin{equation}
{\cal A}_{\lambda,\psi }(z) \rightarrow \frac{1}{z} \,
\frac{1}{(\lambda-\psi)^2} \label{condition}
\end{equation}

The solution of the boundary problem is straightforward. We use the
previous solution\cite{Mig} of the homogeneous problem
\begin{equation}
   G_{\lambda}(\mu) =
\frac{1}{ \lambda-R(\mu)} \Re\ {\cal
T}_{\lambda}(\mu+\imath 0)
\end{equation}
\begin{equation}
\frac{{\cal T}_{\lambda}(\nu+\imath 0)}{{\cal T}_{\lambda}(\nu-\imath 0)} =
\frac{\lambda-R(\nu)+ \imath \pi \rho(\nu)}{\lambda-R(\nu)-\imath \pi
\rho(\nu)}
\label{Bproblem}
\end{equation}
\begin{equation}
{\cal T}_{\lambda}(z) = \exp \left( \int \frac{d \nu}{ \pi (\nu-
z )} \arctan \frac{\pi \rho(\nu)}{\lambda-R(\nu)}\right)
\label{Solution}
\end{equation}
and we find
\begin{equation}
{\cal A}_{\lambda,\psi}(z) = {\cal T}_{\lambda}(z) {\cal
B}_{\lambda,\psi}(z) 
\end{equation}
where the new function $ {\cal B}_{\lambda,\psi}(z) $ which decreases at
infinity precisely as $ {\cal A}_{\lambda,\psi}(z) $ is to be
reconstructed from imaginary part
\begin{equation}
\Im\,{\cal B}_{\lambda,\psi}(z+\imath 0)
= \frac{-  \pi\rho(z) G_{\lambda}(z) \eta(\psi,z)} {{\cal
T}_{\lambda}(z+\imath 0)\left(\lambda -R(z) -\imath \pi \rho(z)
\right)} 
\end{equation}

This expression  can be simplified in virtue of above
equations  as follows
\begin{equation}
\Im\,{\cal B}_{\lambda,\psi}(z+\imath 0)
= \frac{- \pi  \rho(z) \eta(\psi,z)}
 {(\lambda-R(z))^2 + \pi^2\rho^2(z)}
\end{equation}
so that the resulting Cauchy integral reads
\begin{equation}
{\cal B}_{\lambda,\psi}(z) = \int d\nu\,\frac{\rho(\nu) \eta(\psi,\nu)}
{(\lambda-R(z))^2 + \pi^2\rho^2(z)} \,
\frac{1}{z-\nu} 
\end{equation}

Finally, comparing the asymptotic behavior of this integral with known
asymptotics $ {\cal T}_{\lambda}( \infty) =1 $ and required condition
(\ref{condition}) for $ {\cal A}_{\lambda,\psi}(z) $ we find linear
integral equation for $ \eta(\psi,\phi) $
\begin{equation}
\int d\phi\,\frac{\rho(\phi) \eta(\psi,\phi)}
 {(\lambda-R(\phi))^2 + \pi^2\rho^2(\phi)} =
\frac{1}{(\lambda-\psi)^2} 
\end{equation}

Let us now turn to the second $ \phi $ derivative, and differentiate
the above equations for the first derivative. We shall use the same
notation $ g_{\lambda}(\phi',\phi) $ for derivatives of the $ G $
function
\begin{equation}
\sigma(\phi',\phi) = \frac{d}{d \phi'} \frac{\delta
F(\phi)}{\delta \rho_1(\phi')}\\;\;
g_{\lambda}(\phi',\phi) = \frac{d}{d \phi'} \frac{\delta
G_{\lambda}(\phi)}{\delta \rho_1(\phi')}
\end{equation}

The differentiated equations read
\begin{equation}
\int d \mu \rho(\mu)
\frac{g_{\lambda}(\phi',\mu)-g_{\lambda}(\phi',\phi)}{-\mu+\phi} +
(\lambda -F(\phi)) g_{\lambda}(\phi',\phi) =
\sigma(\phi',\phi)G_{\lambda}(\phi) + \frac{d}{d \phi'}
\frac{G_{\lambda}(\phi')-G_{\lambda}(\phi)}{(\phi'-\phi)}
\label{gequation1}
\end{equation}
\begin{equation}
\int d\mu \rho(\mu) g_{\lambda}(\phi',\mu) = 0 \label{extra1}
\end{equation}

This equation differs from the previous one only by the right side.
Repeating the same steps as before we find
\begin{equation}
\int d \mu \frac{\rho(\mu)g_{\lambda}(\phi',\mu)}{z-\mu} ={\cal
T}_{\lambda}(z) \int d \phi \,\frac{\rho(\phi)}{z-\phi}\,
\frac{
	\left(
\sigma(\phi',\phi) +   \frac{1}{G_{\lambda}(\phi)}\frac{d}{d \phi'}
\frac{G_{\lambda}(\phi')-G_{\lambda}(\phi)}{(\phi'-\phi)}
	\right)
}{(\lambda-R(\phi))^2 + \pi^2\rho^2(\phi)}
\end{equation}

Finally, setting $ z \rightarrow \infty $ in above formula and using
the equation (\ref{extra1}) we find equation for $ \sigma(\phi',\phi)
$
\begin{equation}
0 = \int d \phi \,
\frac{\rho(\phi)
	\left(
\sigma(\phi',\phi) +   \frac{1}{G_{\lambda}(\phi)}\frac{d}{d \phi'}
\frac{G_{\lambda}(\phi')-G_{\lambda}(\phi)}{(\phi'-\phi)}
	\right)
}{(\lambda-R(\phi))^2 + \pi^2\rho^2(\phi)}
\end{equation}

\end{document}